# Comparisons of Two Reduced-Order Models for Linearized Unsteady Aerodynamic Identification

Jiaqing Kou,[1] and Weiwei Zhang[2]

*Northwestern Polytechnical University, 710072 Xi'an, People's Republic of China*

**Abstract**: This paper compares the performance of two unsteady aerodynamic reduced-order models (ROMs), namely linear Volterra series and the autoregressive with exogenous input (ARX) model, on modeling dynamically linear aerodynamic behaviors. The difference between these two methods is that the latter model has an autoregressive term while the former model has only the input-related term. The first system is a plunging cylinder in a low-Reynolds number flow, where the flow stable (*Re* < 47). Although the training data can be fitted well with both methods, the linear Volterra method requires a higher model order than the ARX model for the same accuracy. Comparison of the frequency response indicates that the ARX model approximates the frequency response more closely, while the frequency response at high Reynolds number is over-fitted by Volterra series. The second aerodynamic system is a flow over a pitching NACA0012 airfoil, including subsonic and transonic states. The convergence of the model with respect to delay orders, at different Mach numbers and mean angles of attack, is studied in detail. As the Mach number or the mean angle of attack increases, the required delay order will increase. But the ARX model still models this system with a small number of terms at the same level of accuracy. All results indicate that the ARX model outperforms the linear Volterra series in most of cases, especially when the flow is close to the unstable state.

**Keywords**: reduced-order model, Volterra series, autoregressive with exogenous input model, unsteady aerodynamics, linear system, low-Reynolds number flow, transonic flow

---

[1] Graduate Student, School of Aeronautics, National Key Laboratory of Aerodynamic Design and Research; koujiaqing93@163.com

[2] Professor, School of Aeronautics, National Key Laboratory of Aerodynamic Design and Research; aeroelastic@nwpu.edu.cn (Corresponding Author)

## Nomenclature

| | | |
|---|---|---|
| $a, b$ | = | unknown parameters of the autoregressive with exogenous input model |
| $a_\infty$ | = | speed of sound (m/s) |
| $C$ | = | reference length (m) |
| $C_l$ | = | lift coefficient |
| $f$ | = | plunging frequency (Hz) |
| $H$ | = | the Volterra operator |
| $h$ | = | the plunging amplitude (m) |
| $N$ | = | number of samples |
| $m$ | = | delay order of the first-order Volterra series |
| $n, k$ | = | current and delayed time step in discrete time domain |
| $k_f$ | = | reduced frequency, defined as $k_f = \omega C / 2V_\infty$ |
| $p, q$ | = | input and output delay orders of the autoregressive with exogenous input model |
| $R$ | = | radius of the circular cylinder (m) |
| $r$ | = | relative error |
| $St$ | = | Strouhal number |
| $T, T_{real}$ | = | non-dimensional and real time step (s) |
| $t, \tau$ | = | current and delayed time step in continuous time domain |
| $u$ | = | system input |
| $V_\infty$ | = | the freestream velocity (m/s) |
| $y$ | = | system output |
| $\alpha$ | = | angle of attack (degree) |
| $\omega$ | = | the circular frequency of the harmonic motion (rad/s) |

Ⅰ. **Introduction**

In the recent decades, reduced-order models (ROMs)[1, 2] have been greatly developed for aerodynamic modeling, in order to reduce the computational burden of computational fluid dynamics (CFD) in engineering design and characteristic analysis. When a small dynamic perturbation about the steady flow state is considered, all the flow variables vary in a linear fashion. Under this circumstance, the nonlinear, unsteady aerodynamics can be linearized, and one can use a time-linearized or dynamically linear (but statically nonlinear) ROM to describe the aerodynamic behavior. Among these linearized aerodynamic ROMs, linear Volterra series and the autoregressive with exogenous input (ARX) model, are both widely utilized.

Volterra models represent nonlinear systems by a functional infinite series, which was first proposed by mathematician Vito Volterra[3]. After being introduced to nonlinear engineering problems by Wiener[4], the Volterra theory has been widely used in biological engineering[5], system control[6], pattern recognition[7], fault diagnosis[8], etc. Cheng et. al[9] gives a review on applications of Volterra series in engineering problems. In recent decades, this model type is extensively applied to aerodynamic and aeroelastic problems, led by Tromp and Jenkins[10] and Silva[11] in identifying subsonic and transonic aerodynamic loads. These Volterra-based ROMs are now efficient analytical tools for fast aeroelastic analysis and aerodynamic prediction. Applications of this theory include transonic aerodynamic simulation[12], uncertainty qualification[13], aeroelastic analysis[14, 15], ROM across multiple flight conditions[16, 17], limit-cycle oscillation prediction[18], et al.

Although higher order Volterra models are validated to capture nonlinear aerodynamic behaviors accurately, linear Volterra models are of great interest in flutter problems. This is because the analysis of flutter is typically a linear stability problem based on a small disturbance assumption, where a linear aerodynamic representation is needed. However, the linear Volterra model is typically different from some other linear models, like the ARX model, which has been validated efficiently in characteristic analysis of some fluid-structure interaction problems[19, 20]. As indicated by Doyle et al.[6], the general Volterra model belongs to the large class of finite-dimensional nonlinear moving average with exogenous input (NMAX) models, whereas linear Volterra models can be regarded as a type of linear moving average with exogenous input (MAX) model. Compared with nonlinear auto-regressive moving average with exogenous input (NARMAX) models[21] or their linear counterparts of the ARX model, auto-regressive terms are omitted. The ignorance of autoregressive effects may lead to redundancy in model parameters[22, 23]. Besides, from the perspective of intrinsic unsteady flow dynamics, as the fluid damping approaches zero, the influence of previous output becomes dominant, which makes such ignorance questionable. Therefore, it is important to test and compare the performance of Volterra series and the ARX model in modeling such kind of flow phenomenon.

In the present paper, the linear Volterra model and the ARX model are compared, in order to show their applicability in modeling linear aerodynamic characteristics. Both stable incompressible and compressible flows are identified, along with detailed comparison and analysis on the performance of describing different dynamically linear behaviors.

## Ⅱ. Linear Volterra series and ARX models

### A. Volterra Series

Volterra models are capable of describing casual, time-invariant, finite-memory systems. For a single-input/single-output (SISO) system in the continuous time domain, the general system equation described in Volterra series can be shown as

$$y(t) = H_0 + \int_0^t H_1(t-\tau)u(\tau)d\tau$$
$$+ \int_0^t \int_0^t H_2(t-\tau_1, t-\tau_2)u(\tau_1)u(\tau_2)d\tau_1 d\tau_2 + ... \qquad (1)$$
$$+ \int_0^t ... \int_0^t H_s(t-\tau_1, ..., t-\tau_s) \prod_{i=1}^s \{u(\tau_i)d\tau_i\} + ...$$

where $u$ and $y$ represent the input and output of the system, and $t$ represents the current time instant. $H_s$ is the $s$th-order Volterra operator, which is denoted as an $s$-fold convolution between the input and the $s$th-order Volterra operator $H_s$. The first term $H_0$ is the steady-state term. This equation can also be transformed into discrete-time domain

$$y(n) = H_0 + \sum_{k=0}^n H_1(n-k)u(k)$$
$$+ \sum_{k_1=0}^n \sum_{k_2=0}^n H_2(n-k_1, n-k_2)u(k_1)u(k_2) + ... \qquad (2)$$
$$+ \sum_{k_1=0}^n ... \sum_{k_s=0}^n H_s(n-k_1, ..., n-k_s) \prod_{i=1}^s u(k_i) + ...$$

where $H_s$ is now the $s$th-order discrete-time Volterra operator. It should be noted that the discrete-time response is more suitable for numerical and experimental applications, since the continuous-time response is defined in an ideal condition where the input amplitude approaches infinity while its width is zero[15]. For nonlinear system identification, time cost of obtaining high-order Volterra kernel increases exponentially with order. Hence, truncated Volterra series with a low order[12] and pruned high-order Volterra series[24] are usually adopted. When this model is used for dynamic linear systems, e.g., aerodynamic system under small disturbances, only first-order Volterra kernel is needed. Therefore, a truncated, first-order Volterra series is obtained

$$y(n) = H_0 + \sum_{k=0}^n H_1(n-k)u(k) \qquad (3)$$

This model has a typical MAX structure, which calculates current system output based on the discrete-time convolution of the first-order kernel with the system input between time step 0 and $n$. When the linear system is unstable, the kernel will diverge with time and becomes non-zero at infinite time. However, it has to be truncated for practical use and therefore the instability is only partially reflected. To resolve this problem, eigensystem realization algorithm is always used to model the evolution of kernels for unstable systems[25]. For stable systems, to identify the kernel $H_1$, three methods can be adopted, including impulse response, step response and pseudo-inverse methods.

1) *Impulse response*: In a discrete-time system, an impulse signal is defined as

$$u(k) = \begin{cases} \xi, k = 0 \\ 0, k > 0 \end{cases} \tag{4}$$

where $\xi$ refers to a non-zero value for impulse excitation. Response of the system (3) is

$$y(n) = H_0 + \xi H_1(n) \tag{5}$$

Therefore, we have the first-order Volterra kernel:

$$H_1(n) = \frac{y(n) - H_0}{\xi} \tag{6}$$

2) *Step response*: This approach requires two step inputs at time step 0 and 1, respectively. When the step starts at time step 0, the input and the response are defined as

$$u(k) = \xi, k \geq 0 \tag{7}$$

$$y_0(n) = H_0 + \xi \sum_{k=0}^{n} H_1(n - k) \tag{8}$$

The corresponding input and response at time step 1 are

$$u(k) = \begin{cases} 0, k = 0 \\ \xi, k \geq 1 \end{cases} \tag{9}$$

$$y_1(n) = H_0 + \xi \sum_{k=1}^{n} H_1(n - k) \tag{10}$$

From (8) and (10), the first-order Volterra kernel is identified as:

$$H_1(n) = \frac{y_0(n) - y_1(n)}{\xi} \tag{11}$$

This algorithm is also used for identifying linear step-type ROM described by indicial functions, since the first-order Volterra system and the linear step-type ROM is identical[26]. Compared with the impulse-type Volterra series, step-type based models are more robust to different input amplitudes and time steps[12].

3) *Pseudo-inverse method*: Since identifying discrete-time Volterra series from training signals can be regarded as solving a set of linear equations, standard least-squares approach like pseudo-inverses can be used. This approach not only gives more robust and accurate ROMs, but also allows random training signals[18]. Consider $N+1$ input-output samples from full-order simulation, denoted as $\boldsymbol{u} = [u(0),...,u(N)]^T$ and $\boldsymbol{y} = [y(0)-H_0,...,y(N)-H_0]^T$, respectively. The unknown first-order Volterra series is calculated as:

$$\boldsymbol{H}_1 = \boldsymbol{M}^+ \boldsymbol{y} \tag{12}$$

where + is the Moore–Penrose pseudo-inverse. $\boldsymbol{H}_1$ is the vector containing all terms of the first-order Volterra kernel, i.e., $\boldsymbol{H}_1 = [H_1(m), H_1(m-1), ..., H_1(0)]^T$, where $m$ indicates the memory (delay order) of the kernel. The matrix $\boldsymbol{M}$ has the inputs at different time steps, defined as

$$\boldsymbol{M} = \begin{bmatrix} u(0) & 0 & \cdots & 0 \\ u(1) & u(0) & \cdots & 0 \\ \vdots & \vdots & \ddots & \vdots \\ u(N) & u(N-1) & \cdots & u(N-m) \end{bmatrix} \tag{13}$$

For an accurate identification of Volterra kernels, (12) should be a well-posed problem. Hence, filtered white Gaussian noise (FWGN) signal, which is a random signal with wide range of amplitudes and frequencies, can be used for model training. This will help to avoid linear dependence of matrix $\boldsymbol{M}$. In the current study, all the three methods will be compared firstly and the pseudo-inverse method is then used for accessing the performance of linear Volterra series.

**B. ARX model**

The ARX model represents linear dynamic system by a discrete-time difference equation:

$$y(n) = \sum_{i=1}^{q} a_i y(n-i) + \sum_{i=0}^{p-1} b_i u(n-i) \tag{14}$$

Compared with first-order Volterra model in (3), this model has autoregressive effects, expressed by output time-delayed terms. Using the backward shift operator $z$, i.e., $y(n-1) = z^{-1} y(n)$, this system

can be further denoted as

$$y(n) = \frac{B(z)}{A(z)} u(n) \qquad (15)$$

where $A(z)$ and $B(z)$ are polynomials in the backward shift operator. The poles and zeros of the system are shown in $A(z)$ and $B(z)$, respectively. Based on (15), it can be concluded that for linear Volterra series, $A(z) = 1$. Therefore, all of the poles of the linear Volterra system are at zero[27]. So this model may fail to capture the dynamics if a system has a non-zero pole point. The ARX model is easily identified by least-squares method. Define vector $\boldsymbol{\theta}$ of all unknown parameters, i.e., $\boldsymbol{\theta} = \begin{bmatrix} a_1 & \cdots & a_q & b_0 & \cdots & b_{p-1} \end{bmatrix}^T$, (15) can be written in the vector form

$$\begin{cases} y(n) = \boldsymbol{\theta}^T \boldsymbol{x}(n) \\ \boldsymbol{x}(n) = \begin{bmatrix} y(n-1) & \cdots & y(n-q) & u(n) & \cdots & u(n-p+1) \end{bmatrix}^T \end{cases} \qquad (16)$$

where $\boldsymbol{x}(n)$ is the state vector. The estimated parameter vector $\hat{\boldsymbol{\theta}}$ is calculated as follows:

$$\begin{cases} \hat{\boldsymbol{\theta}} = (\boldsymbol{X}^T \boldsymbol{X})^{-1} \boldsymbol{X}^T \boldsymbol{y} \\ \boldsymbol{X} = \begin{bmatrix} \boldsymbol{x}^T(1) \\ \boldsymbol{x}^T(2) \\ \vdots \\ \boldsymbol{x}^T(N) \end{bmatrix} = \begin{bmatrix} y(0) & \cdots & y(1-q) & u(1) & \cdots & u(2-p) \\ y(1) & \cdots & y(2-q) & u(2) & \cdots & u(3-p) \\ \vdots & & \vdots & \vdots & & \vdots \\ y(N-1) & \cdots & y(N-q) & u(N) & \cdots & u(N+1-p) \end{bmatrix} \end{cases} \qquad (18)$$

### III. Results and Discussions

In order to show the applicability of linear Volterra model in identifying linear aerodynamics, two typical flow conditions from numerical simulation are chosen: 1) Flow past a circular cylinder at low Reynolds numbers; 2) Subsonic and transonic flow past an airfoil at different angles of attack. Note that in both test cases, the system is excited with a small disturbance to ensure linear aerodynamic behaviors, and all the considered aerodynamic systems are stable. An in-house computational fluid dynamics (CFD) code based on hybrid unstructured mesh is used to achieve accurate numerical simulation. This code solves the unsteady laminar Navier-Stokes equations or unsteady Reynolds-averaged Navier–Stokes (URANS) equations using a cell-centered finite volume approach.

For flow past a cylinder at low Reynolds numbers, the incompressible laminar flow is simulated at Mach number 0.1. For simulation of flow past an airfoil, URANS equations with the S-A turbulence model [28] are solved. Time scale is non-dimensionalized by speed of sound $a_\infty$ and reference length $C$ (cylinder diameter or chord length of an airfoil), defined as $T = a_\infty T_{real}/C$. Details about numerical methods are given in Zhang et al.[19] and Gao et al.[20, 29].

**A. Flow past a circular cylinder**

The flow past a circular cylinder is a classical problem in the study of fluid dynamics[30]. It is well known that the flow around a stationary cylinder becomes unstable when the Reynolds number is larger than 47, accompanied by the periodic von Kármán vortex shedding phenomenon. However, when $Re < 47$, due to fluid-structure interaction of the elastically supported structure, instability will also occur, which is known as the vortex-induced vibration. To analyze the stability at these subcritical $Re$, an accurate ROM is needed. Therefore, we firstly test the performance of linear Volterra series by modeling the unsteady flow fields at Reynolds numbers 12, 25 and 45, respectively. Both linear Volterra series and the ARX model are used to construct the mapping between plunging motion and lift coefficient. The plunging amplitude is denoted as $h/R$ (positive up, so is lift coefficient), where $R$ is the cylinder radius.

Training signal is very important to ROMs. Similar with Zhang et al.[19], here we choose a chirp signal with a broadband frequency coverage as the training case, as shown in Fig. 1. This signal has 1400 data points and allows a wide range of excitation on the frequencies we need. The moving frequency is characterized by Strouhal number, i.e., $St = \dfrac{2fR}{V_\infty}$, where $f$ is the plunging frequency and $V_\infty$ is the freestream velocity. Here $St_{flow}$ is used to define the characteristic frequency of the fluid system. The intrinsic flow frequency and damping can be determined by many methods like curve fitting and dynamic mode decomposition[31]. Fig. 2 shows the relationship between flow frequency and damping from $Re = 12$ to $Re = 60$. This figure indicates that, as $Re$ increases, the fluid system becomes less stable. Therefore, it is important to investigate how ROM works at different conditions.

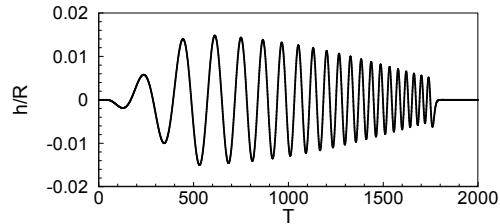

Fig. 1    Training signal for flow past a circular cylinder at low Reynolds numbers.

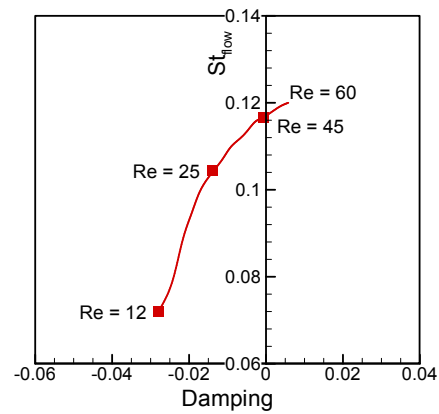

Fig. 2    Strouhal number versus system damping from Reynolds number 12 to 60.

Firstly, three identification methods for Volterra series are compared, as shown in Fig. 3. For illustration purpose, only results at $Re = 12$ is given. Note that the Volterra kernel of each method is calculated by the impulse signal, the step signal and the chirp signal, respectively. And in Fig. 3b, these models are all used to predict the response of the chirp signal. Relative error $r$ is introduced to evaluate the performance of all three ROMs[18].

$$r = \sqrt{\frac{\sum_{i=1}^{N}|y_{ROM}(i)-y_{CFD}(i)|^2}{\sum_{i=1}^{N}|y_{CFD}(i)|^2}} \times 100\% \qquad (19)$$

where $y_{ROM}$ and $y_{CFD}$ are outputs of ROM and CFD solver, respectively. $N$ is the total number of test data. From Fig. 3a, it is evident that identifying Volterra kernels based on pseudo-inverse approach gives better convergence with delay order and lower error at a specific order. The identified aerodynamic responses at delay order $m = 50$, as well as CFD output data, are compared in Fig. 3b, where pseudo-inverse-based Volterra model gives best approximation. Therefore, in the following study, pseudo-inverse identification approach is adopted.

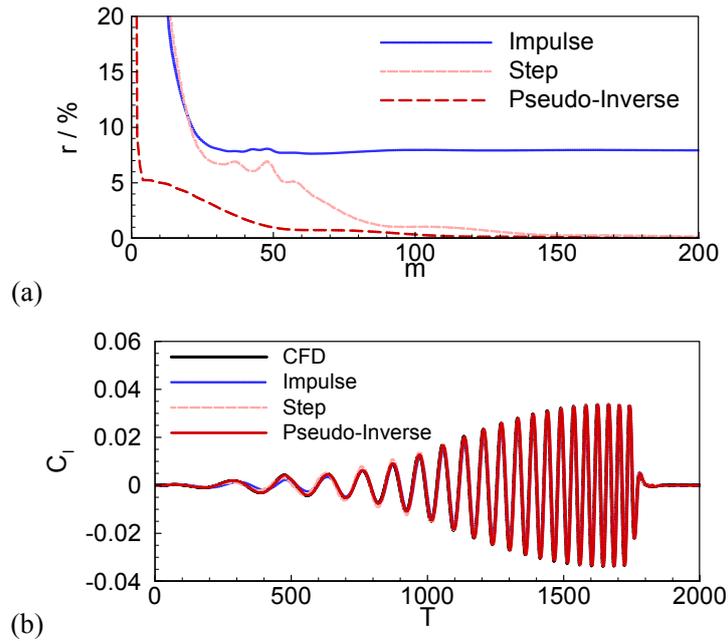

(a)

(b)

**Fig. 3 Comparison of three identification methods at $Re = 12$. a) Relative error versus delay order. b) The predicted responses at $m = 50$.**

The relative error of both ARX and Volterra models, changing with delay orders, are shown in Fig. 4. Note that the delay order of ARX model is the sum of input and output delay orders (assuming $p = q$ for illustration purpose, but in practice they are not required to be equal). These figures indicate that as

*Re* increases, relative error converging with delay order becomes slower. Therefore, a small delay order suitable for *Re* = 12 is not enough for test cases at *Re* = 45. For example, if the user wants to have a linear Volterra ROM within 5% training error, the expected delay orders for Reynolds number 12, 25, 45 are 10, 75 and 740, respectively. So we need to change the delay order according to different flow conditions. From the perspective of flow stability, it is easy to find that, the more unstable flow becomes, the larger delay order is needed. Besides, convergence of ARX model is much faster than linear Volterra model. This makes ARX model have a more parsimonious structure (i.e., a smaller number of unknown parameter) to allow accurate characteristic analysis.

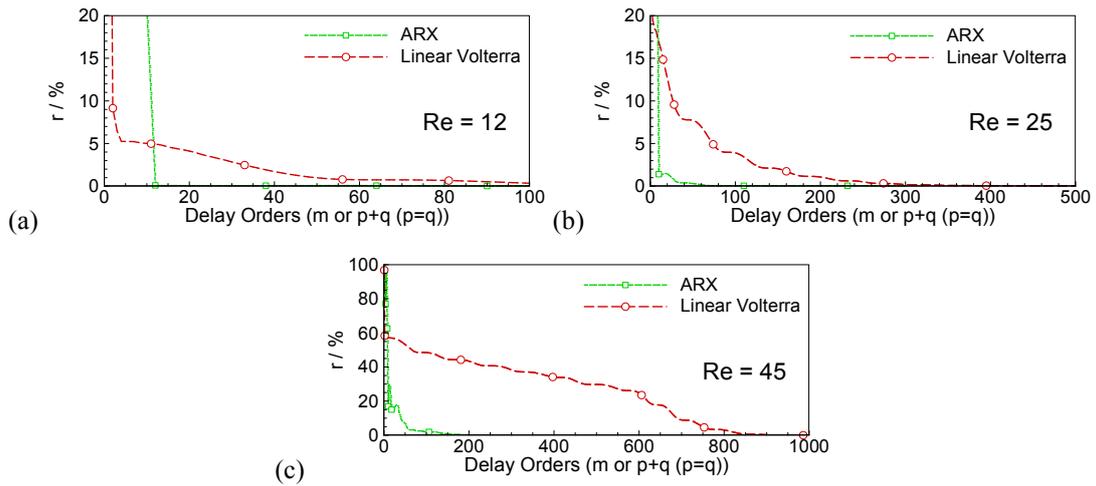

**Fig. 4 Relative error at different delay orders. a) *Re* = 12 b) *Re* = 25 c) *Re* = 45.**

In most of the previous studies[18, 25, 32, 33] on Volterra series, the flow is very stable, whose damping may be smaller than that of flow past a cylinder at *Re* = 12. Therefore, using a delay order $m <$ 100 is enough to provide a good representation of aerodynamics. But as the flow approaches critical stable state (damping increases to zero), this delay order is not enough. This is validated in Fig. 4c, where using a delay order larger than 800 will lead to acceptable accuracy.

When both ROMs have the same training accuracy, it is interesting to see whether they have the same performance in capturing dynamic characteristics at different frequencies. Therefore, now we focus on the frequency response. The selected delay orders are listed in Table I. For both ROMs, the delay order is selected to make the training relative error lower than 1%. Note that for all the test cases, delay orders of the ARX model are much smaller than those of linear Volterra model. Moreover, in order to make an equal comparison between ARX and linear Volterra ROMs, linear Volterra series with the same numbers of delay order are also constructed. We compare each ROM's performance through Bode plots, as shown in Figs. 5-9. The CFD results are given by calculating the frequency response at 6 frequencies: $St/St_{flow}$ = 0.5, 0.9, 1, 1.1, 1.5, 2.

**Table 1 Selected delay orders at different flow conditions.**

| $Re$ | ARX ($p+q$, $p=q$) | Linear Volterra ($m$) |
|---|---|---|
| 12 | 30 | 30, 100 |
| 25 | 120 | 120, 320 |
| 45 | 160 | 160, 900 |

Because we are interested in the ROM performance for unsteady flow at different stability characteristics, three Reynolds number, 12, 25 and 45 are chosen. When the flow is very stable, as shown in Fig. 5, the ARX model with delay order 30 capture the frequency response very well. However, if the linear Volterra ROM with the same delay order is considered, it shows large errors, especially in low-frequency features. The performance of linear Volterra ROM becomes comparable to that of the ARX model when the delay order is set large enough like 100. As $Re$ increases, shown in Fig. 6, the flow damping approaches to zero, and acceptable delay order increases. But the trend is the same, where a Volterra ROM with a much larger delay order shows an agreement with an ARX ROM with a lower order. These two groups of test case show that, with higher delay orders, linear Volterra models are capable of presenting the same frequency responses as those of the ARX model. When the flow is going to be unstable (*Re* = 45), as shown in Fig. 7, it is observed that using a very large delay order (*m* = 900) cannot guarantee good prediction of frequency responses. This is caused by the fact that linear Volterra series does not contain an autoregressive term, which is needed for the proper representation of unsteady aerodynamics. Note that for very stable unsteady flows, Volterra series also works well[25, 34, 35], but it is seldom used when the flow is close to the unstable state. Besides, oscillatory behavior is also observed in Fig. 7, when the frequency response of the Volterra series model with delay order 900 is calculated. This is caused by the overfitting phenomenon since the kernel contains too much underdetermined terms, leading to very large condition number of the matrix $M$ of Eq. (12). This phenomenon is also found in nonlinear ROMs[36]. When an aerodynamic ROM with only input delay orders is introduced, we need a large number of terms to approximate the aerodynamic loads. However, introducing output feedback will lead to much smaller delay orders with improved performance.

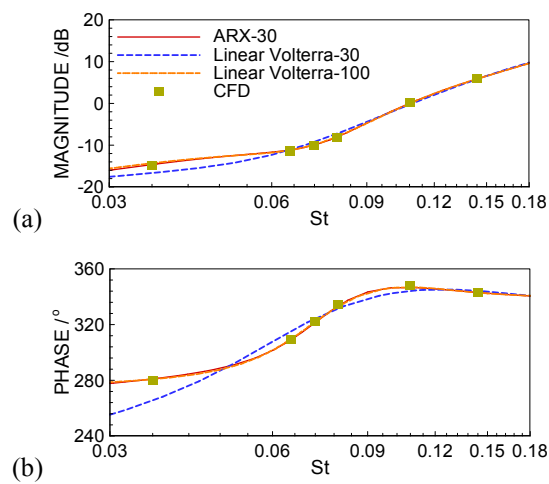

**Fig. 5 Bode plot for ROMs at *Re* = 12. a) Magnitude response b) Phase difference.**

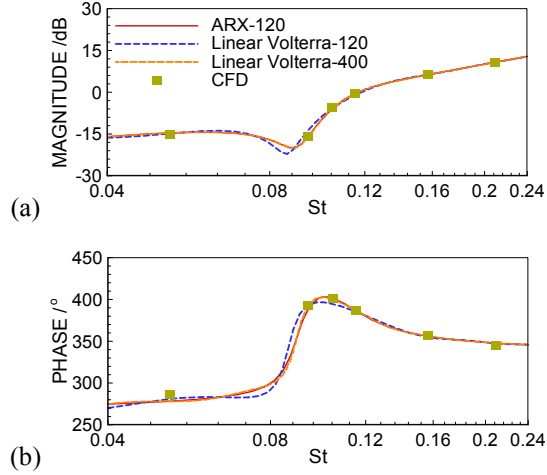

**Fig. 6 Bode plot for ROMs at *Re* = 25. a) Magnitude response b) Phase difference.**

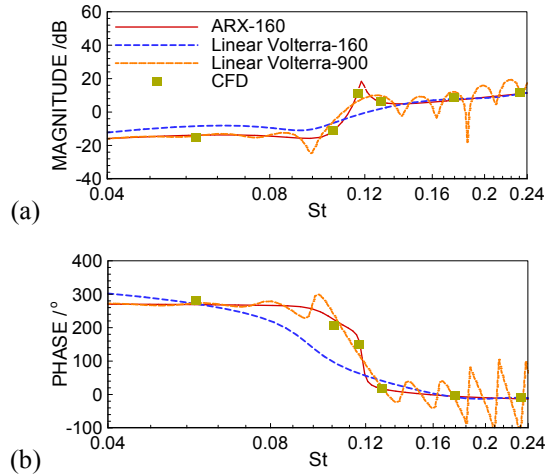

**Fig. 7 Bode plot for ROMs at *Re* = 45. a) Magnitude response b) Phase difference.**

In Figs. 8-9, influence of delay orders is studied at $Re$ = 45. For ARX models, good robustness to delay orders is seen. As long as the delay order meets the requirement of accuracy, indicated by Fig. 4c, the frequency response is captured with a reasonable accuracy. However, linear Volterra models show different frequency responses across a range of delay orders. Even though a large delay order $m = 1100$ is chosen (training error smaller than 0.1%), the predicted frequency response is unsmooth and unreliable. This is also due to the overfitting problem during model identification. The limitation of

Volterra series is also indicated in Fig. 10, where temporal harmonic responses with a small disturbance at different frequencies are shown. Three frequencies, which are close to the characteristic fluid frequency are chosen. Although all models have large errors in predicting frequency response at $St_{flow}$, ARX models outperform linear Volterra models in other cases. At $St/St_{flow} = 1$, resonance is strong (corresponding to the pole point of the system), which is very difficult to reproduce. But this will not influence the overall performance, as shown in Figs. 7-8.

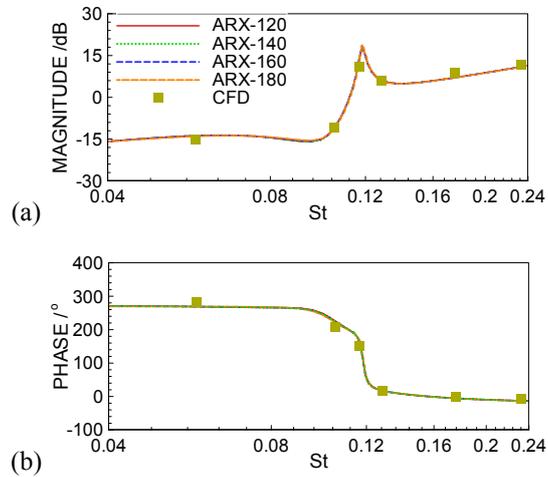

**Fig. 8 Bode plot for ARX-based ROMs at different delay orders and *Re* = 45. a) Magnitude response b) Phase difference.**

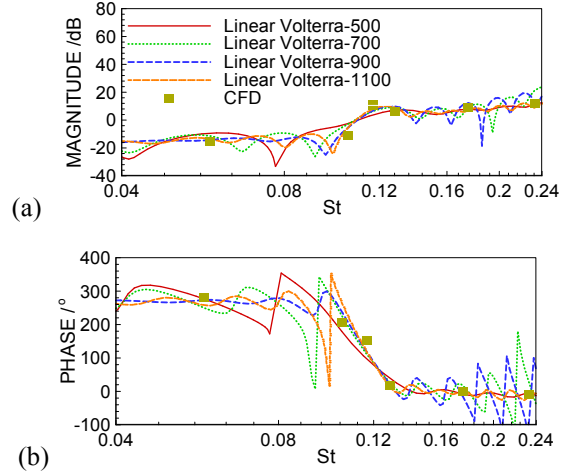

**Fig. 9 Bode plot for linear Volterra ROMs at different delay orders and *Re* = 45. a) Magnitude response b) Phase difference.**

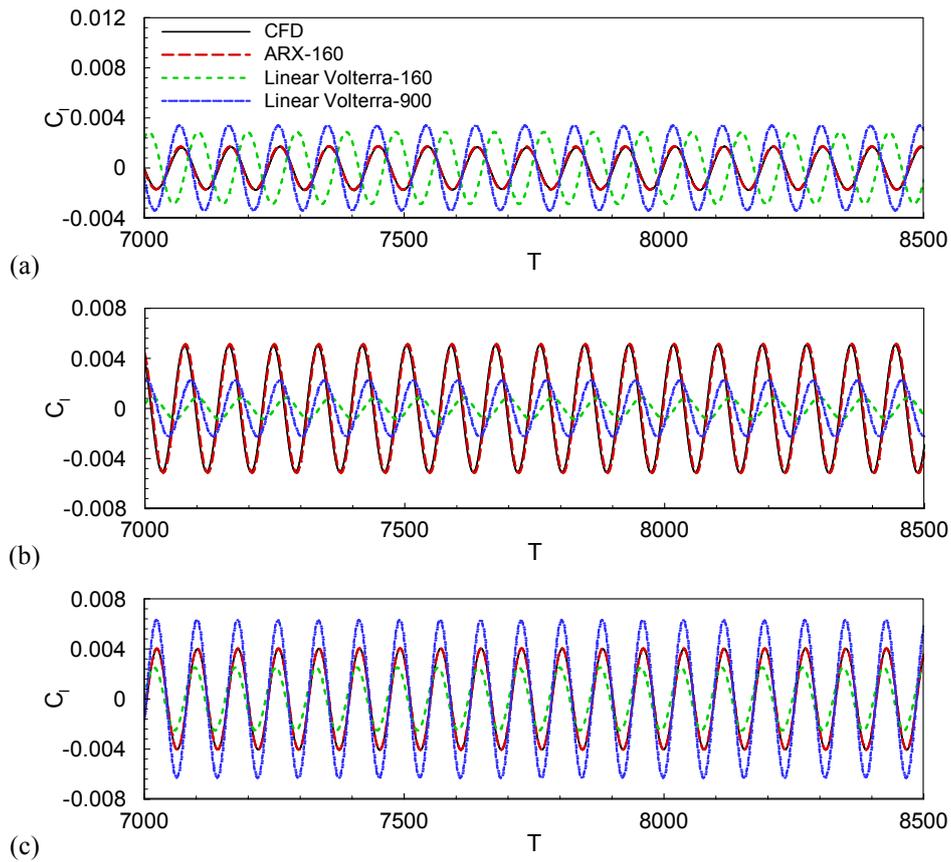

**Fig. 10 Aerodynamic prediction at Reynolds number 45 ($St_{flow}$ =0.1167). a) $h/R$ = 0.006, $St/St_{flow}$ = 0.9. b) $h/R$ = 0.001, $St/St_{flow}$ = 1. c) $h/R$ = 0.002, $St/St_{flow}$ = 1.1.**

**B. NACA0012 airfoil in subsonic and transonic flows**

To further test the performance of these two models in modeling linear dynamic behaviors, the other test case, i.e., flow past a NACA0012 airfoil, is considered. Both models are used to identify the dynamical relationship between the pitching displacement and the lift coefficient. To capture the frequency response accurately, a chirp signal is used for training, as shown in Fig. 11. The airfoil is pitching at the mid-chord point. At mean angle of attack $\alpha_0 = 0°$, five Mach numbers ($Ma$ = 0.50, 0.60, 0.65, 0.70 or 0.75) are selected, ranging from moderate to transonic flow conditions. We give the relationship between the relative error and the delay order, as shown in Fig. 12. In Fig. 12a, some peaks are observed at $Ma$ = 0.70 and $Ma$ = 0.75. Since an autoregressive term is used in the ARX model, the resulting ARX model is dynamic, therefore the model output will also be influenced by previous model outputs. This will explain the occurrence of the peak value because if the model is not properly fitted, errors of previous output will influence those of the future output. But the overall performance of the ARX model will be improved with increasing delay orders. For transonic flow with moving shock waves, this phenomenon may sometimes happen and the user needs to adjust the delay orders. However, it is shown that for all Mach numbers, the relative error decreases fast and when the delay order is larger than 40, very low error ($r \leq 1\%$) is obtained. But for linear Volterra models, required delay order largely relies on flow conditions, as indicated in Fig. 12b. As $Ma$ increases, large delay order is need for a reasonable accuracy. Besides, to reach the same relative error, more terms in Volterra series is needed. For example, at $Ma$ = 0.50, if one wants to obtain a ROM with relative error 1%, delay orders for ARX and Volterra models are 16 and 100, respectively. The difference is even larger for test case at $Ma$ = 0.75, indicating a difficulty in modeling transonic flows from linear Volterra series. A harmonic motion at $Ma$ = 0.75 and reduced frequency 0.2793 is used to test the performance of both models. The reduced frequency is defined as $k_f = \omega C / 2V_\infty$, where $\omega$ is the circular frequency of the harmonic motion. The prediction is shown in Fig. 13, where the ARX model gives the most accurate prediction with only delay order 10. For Volterra series, it exhibits very large errors even though the delay order is set to 40.

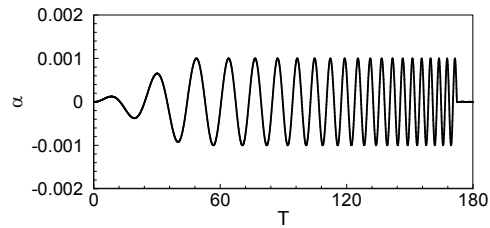

Fig. 11 Training signal for flow past an airfoil.

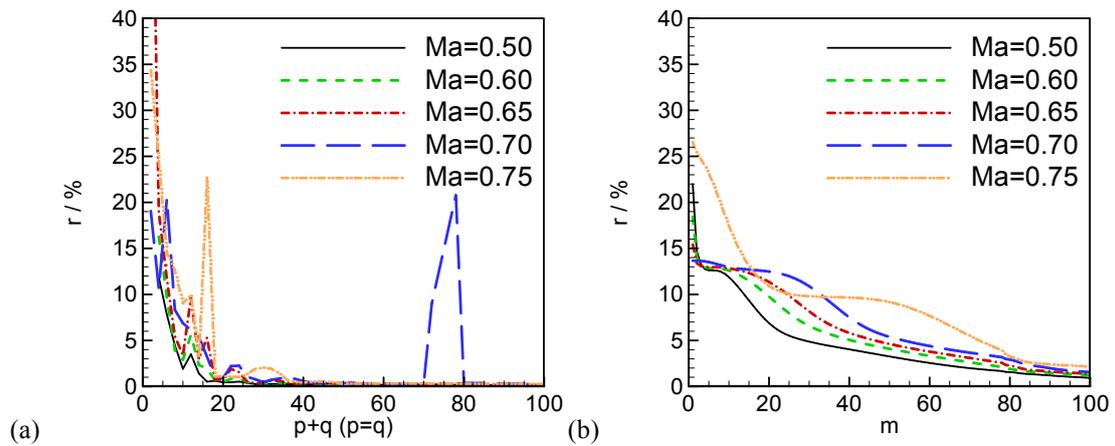

(a) (b)

Fig. 12 Relative error at different delay orders and Mach numbers (zero mean angle of attack). a) ARX model b) Linear Volterra model.

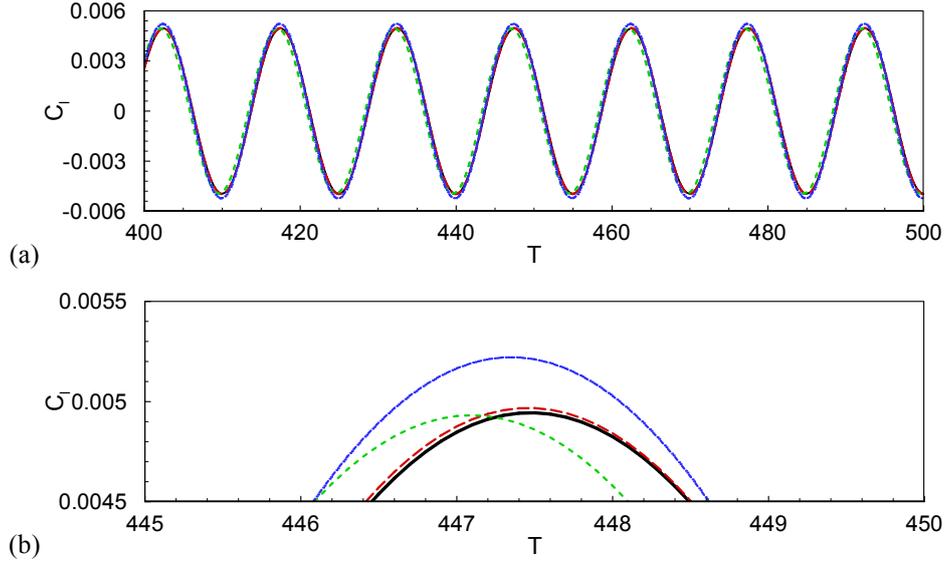

**Fig. 13 Aerodynamic prediction at Mach number 0.75, mean angle of attack 0º, with reduced frequency 0.2793 and amplitude 0.001 rad. Black: CFD. Red: ARX with delay order 10. Green: Linear Volterra with delay order 10. Blue: Linear Volterra with delay order 40.**

Transonic flow is a typical flow condition for ROM-based aeroelastic analysis, so we test both models at $Ma = 0.7$. From Gao et al.[37], the computational buffet onset is $\alpha = 4.8°$, where the flow damping becomes zero, resulting in a limit cycle oscillation with periodically moving shock waves. We choose different mean angles of attack to characterize different stability features. The comparison between both models are shown in Fig. 14. As $\alpha_0$ increases, both models need a larger delay order for acceptable accuracy. However, ARX model still converges faster than Volterra model with delay orders (note that in Fig.13b, $m$ changes from 1 to 1000). For Volterra models, as shown in Fig. 14b, delay order changes greatly as flow approaches unstable. For $\alpha_0 = 4.5°$, only 20% relative error is obtained with 1000 terms. But for ARX model with the same accuracy, 50 terms are needed. The performance of the models for Mach number 0.7 and mean angle of attack 4.3º is compared in Fig. 15, by predicting the aerodynamic coefficient of a harmonic motion with reduced frequency 0.2992 and amplitude 0.001 rad. It is noted that ARX model still shows better accuracy with a small delay order. These results indicate that when we want to model linear systems approaching instability, Volterra series is not a good choice.

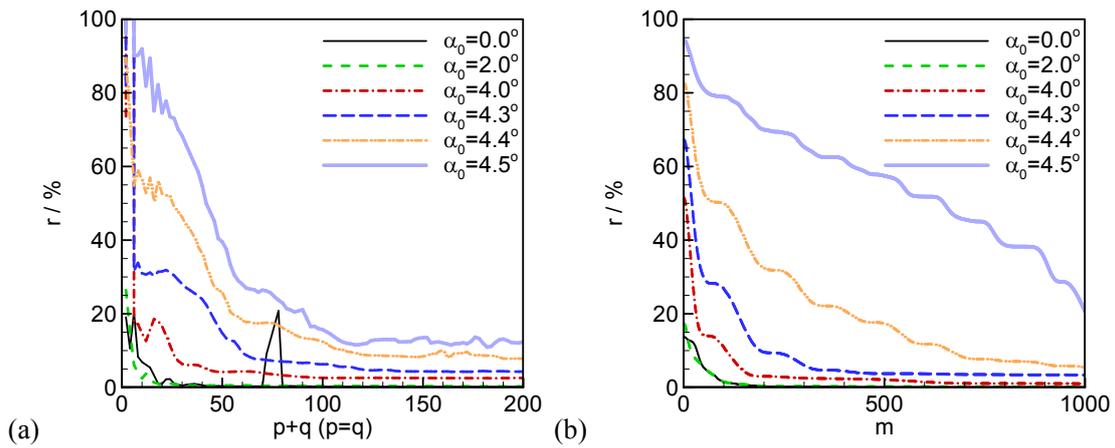

Fig. 14 Relative error at different delay orders and mean angles of attack (Mach number 0.70). a) ARX model b) Linear Volterra model.

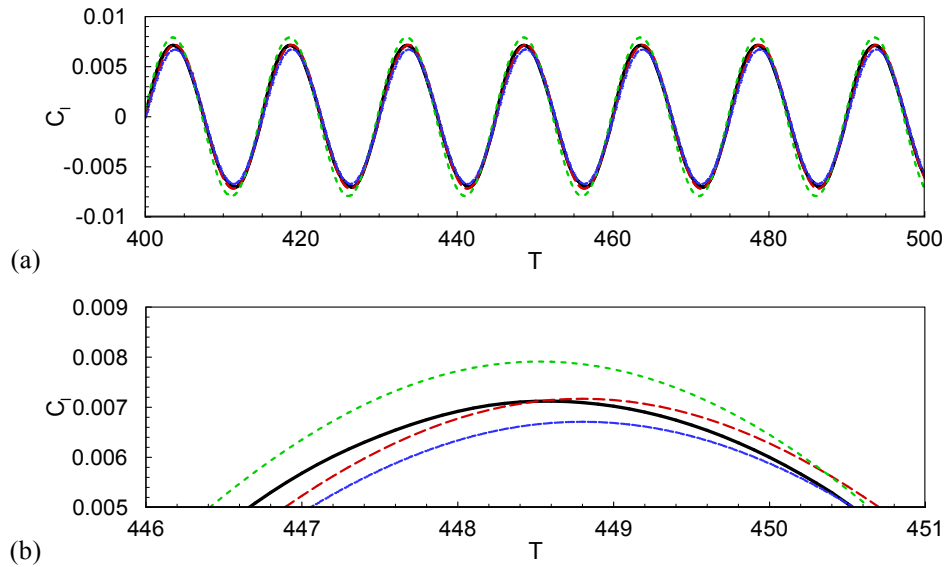

Fig. 15 Aerodynamic prediction at Mach number 0.7, mean angle of attack 4.3º, with reduced frequency 0.2992 and amplitude 0.001 rad. Black: CFD. Red: ARX with delay order 50. Green: Linear Volterra with delay order 50. Blue: Linear Volterra with delay order 200.

## Ⅳ. Conclusions

This study compares the applicability of first-order Volterra series and the ARX model in modeling linearized unsteady aerodynamics. It is found that linear Volterra series is suitable for very stable flows, but the required delay order increases and overall accuracy decreases as the flow approaches instability. However, the ARX model shows better accuracy with small delay orders, and will approximate the frequency responses precisely. This is caused by the fact that the Volterra series does not contain autoregressive terms, therefore the unsteady effects are only partially exhibited through the input-delayed terms. Therefore, one should take care when using Volterra series for linear aerodynamic systems, especially for transonic flows or separated flows approaching an unstable state.

## Acknowledgement

This work was supported by the National Natural Science Foundation of China (No. 11572252), the National Science Fund for Excellent Young Scholars (No. 11622220), 111 project of China (No. B17037) and ATCFD project (2015-F-016).